\documentclass[3p,times,preprint]{elsarticle}

\usepackage{ecrc}

\volume{00}

\firstpage{1}

\runauth{C. Estarellas and L. Serra}

\usepackage{amssymb}

\usepackage[figuresright]{rotating}

\begin{document}

\begin{frontmatter}



\dochead{}

\title{A scattering model of 1D quantum wire regular polygons}


\author[ifisc]{Cristian Estarellas}
\author[ifisc,dep]{Lloren\c{c} Serra\corref{cor1}}
\cortext[cor1]{Corresponding author.}
\ead{llorens.serra@uib.es}

%
%
\address[ifisc]{
Institut de F\'{\i}sica Interdisciplin\`aria i de Sistemes Complexos
IFISC (CSIC-UIB), E-07122 Palma de Mallorca, Spain}
\address[dep]{
Departament de F\'{\i}sica,
Universitat de les Illes Balears, E-07122 Palma de Mallorca, Spain}

\begin{abstract}
We calculate the quantum states of regular polygons made of 1D quantum wires
treating each polygon vertex as a scatterer. The vertex scattering matrix 
is analytically obtained from the model of a circular bend of a given angle
of a 2D nanowire. In the single mode limit the spectrum is classified in 
doublets 
of vanishing circulation, twofold split by the small vertex reflection, and 
singlets 
with circulation degeneracy. Simple analytic expressions of the energy eigenvalues 
are given. It is shown how each polygon is characterized by a specific spectrum.
\end{abstract}


\end{frontmatter}


\section{Introduction}
\label{intro}

Nanowires are a long-lasting topic of interest in nanoscience
for two main reasons.  
One is, undoubtedly, their use as electron waveguides in devices and 
technological applications.
The other is the possibility of using nanowires to artificially control quantum properties
at the nanoscale, thus improving our fundamental understanding and prediction
capabilities. In nanowires made with 2D electron gases of semiconductors like GaAs 
quantum behavior manifests at a mesoscopic scale, beyond the atomistic description, and it can be described with 
effective mass models and smooth potentials \cite{Datta,Ferry}.

\begin{figure}[t]
\centering
\resizebox{0.35\textwidth}{!}{
	\includegraphics{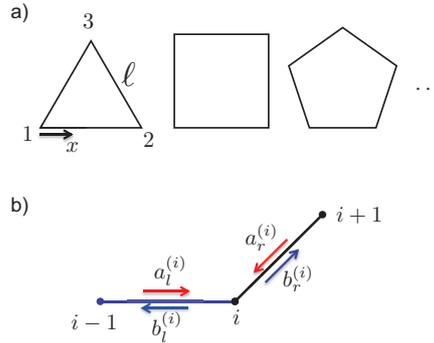}
}
\caption{
a) Sketch of the regular polygons made of 1D nanowires. The vertex labeling
and the definition of a longitudinal coordinate $x$ is indicated only for the triangle.
b) Scattering amplitudes as defined for a given vertex $i$.
}
\label{F1}
\end{figure}

The quantum states on nanowire bends attracted much interest some years 
ago \cite{Lent,Sols,Sprung,Wu,Wu2,Wu3,Vacek,Xu}.
It was shown that bound states form on the bend, and the transmission and
reflection properties as a function of the energy 
and width of the wire
were 
thoroughly
investigated.
In the single-mode limit of a 2D wire a bend can be described with an
effective 1D model containing an energy-dependent potential \cite{Wu}. 
In the case of a circular
bend, the situation becomes simpler and analytical approximations using a square well
whose depth and length are fixed by the radius and angle of the bend were 
suggested \cite{Sprung}.

Motivated by the above mentioned interest on nanowire bends we study in this work
polygonal structures made of 1D quantum wires (Fig.\ \ref{F1}). We focus on the single-mode limit
and describe each vertex as a scatterer. The vertex scattering matrix is taken from the
model of circular bends in 2D nanowires. We find that the spectrum fully characterizes 
the polygon structure. Two kinds of states are present: a) doublets, with a small energy 
splitting due to reflection and with a vanishing circulation along the perimeter of the 
polygon; and b) singlets, with an underlying degeneracy by which the circulation can
take arbitrary values within a given range.
The energy splitting of the doublets depends on the reflection probability
of the vertex, while the number of singlets between consecutive doublets (in energy)
can be used to infer the number of vertices. The energy scale, setting  the separation between 
states, is fixed by the length of the nanowire  forming the sides of the polygon. 

We finally remark that, although 
our approach is on physical modeling of nanostructures, there is a connection with the
more mathematically oriented study of quantum graphs \cite{Berko}. 
Indeed, from this perspective our results 
can be viewed as a particular application to the case of regular polygons of the more general 
question of reconstructing the graph topology from the knowledge of the spectrum.

\section{Theoretical model and method}
\label{theory}

We consider a regular polygon with $N_v$ vertices, 
made of 1D quantum wires of length $\ell$ (Fig.\ \ref{F1}). 
The distance along the perimeter is measured by variable $x$, 
with origin arbitrarily taken on the first vertex.
The wave function between vertices $i$ and $i+1$ is a superposition 
of left- and right-ward propagating plane waves
\begin{eqnarray}
\psi(x)&=& 
a_r^{(i)} e^{-ip(x-x_i)}+
b_r^{(i)} e^{ip(x-x_i)}\nonumber\\
&=&
a_l^{(i+1)} e^{ip(x-x_{i+1})}+
b_l^{(i+1)} e^{-ip(x-x_{i+1})}\; ,
\end{eqnarray}
where the first and second equalities take the reference point 
on vertices $i$ and $i+1$, respectively. The $a^{(i)}_{l/r}$ and $b^{(i)}_{l/r}$ coefficients
are the characteristic input and output scattering amplitudes 
defined in Fig.\ \ref{F1}b. We are assuming a single propagating 
mode of wavenumber $p$ (see \ref{apA}) and the set of vertex positions is
 $\{ x_i, i=1,\dots,N_v\}$.

The $i$-th vertex scattering equation relates output and input 
amplitudes as
\begin{equation}
\label{ivs}
\left(
\begin{array}{c}
 b_l^{(i)}\\
\rule{0cm}{0.5cm} b_r^{(i)}
\end{array}
\right)
=
\left(
\begin{array}{cc}
  r & t \\
 t & r 
\end{array}
\right)
\left(
\begin{array}{c}
  a_l^{(i)}\\
\rule{0cm}{0.5cm}  a_r^{(i)}
\end{array}
\right)\; ,
\end{equation}
where $t$ and $r$ are the usual transmission and reflection complex coefficients.
The scattering matrix in Eq.\ (\ref{ivs}) is summarizing the physical
effect of the vertex by means of two complex quantities, $t$ and $r$.
As they are required inputs in our model, we will take these values
from the known scattering matrices of circular bends in 2D nanowires.
Actually, as discussed in \ref{apA}, in the single-mode limit there is an analytical description of the
scattering by a circular bend in terms of an effective square well
depending on the radius and angle of the bend \cite{Sprung}.

There is a relation between input and output amplitudes of successive vertices,
\begin{eqnarray}
\label{bli}
b_l^{(i)} &=& a_r^{(i-1)} e^{-ip\ell}\; ,\\
\label{bri}
b_r^{(i)} &=& a_l^{(i+1)} e^{-ip\ell}\; .
\end{eqnarray}
The phase  $e^{-ip\ell}$ in Eqs.\ (\ref{bli}) and (\ref{bri}) appears due to
the assumption of two different reference points when considering the 
scattering processes from two consecutive vertices.

We define the circulation
\begin{equation}
\label{circ}
{\cal C} \equiv \frac{\hbar p}{m} 
\left(\left| a_l^{(i)}\right|^2-\left|b_l^{(i)}\right|^2\right)
N_v \ell \; ,
\end{equation}
independent on the choice of vertex $i$ due to flux conservation. The circulation 
is also the same if one chooses the right coefficients ($a^{(i)}_r$ and $b^{(i)}_r$) instead of the left
ones in Eq.\ (\ref{circ}). Physically, the circulation is measuring
the probability current flowing along the polygon in a particular state,
characterized by a set of $a$ and $b$ coefficients and a wave number $p$ 
(corresponding to an energy $E$). 

In \ref{apB}  we define an energy-dependent measure ${\cal F}(E)$,
such that it vanishes for the eigenenergies of the polygon. In practice, we scan 
numerically the values of ${\cal F}(E)$ in order to determine the eigenenergies
within a given energy interval.

\section{Results}

\begin{figure}[t]
\centering
\resizebox{0.48\textwidth}{!}{
	\includegraphics{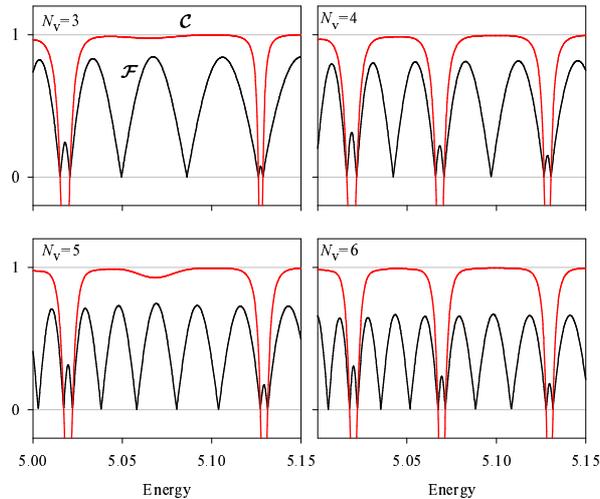}
}
\caption{
Energy dependence of the 
$\mathcal{F}$ measure in arbitrary units (black)
and the circulation ${\cal C}$ in units of $\hbar p /m$ (gray-red).
The unit of energy is $\hbar^2/md^2$.
The wave number $p$ and the wire parameter $d$ are defined in Sec.\ \ref{theory}.
As indicated, each panel corresponds to a polygon with a different number of vertices $N_v$ but the same arm length $\ell=30 d$.
The figure shows the presence of 
doublets with vanishing circulation and singlets with maximal positive 
circulation ${\cal C}_0$. Notice that at the energies of the singlets the circulation
can take values from $-{\cal C}_0$ to $+{\cal C}_0$.
Other parameters: $R=1.3d$.
}
\label{FAA}
\end{figure}

Figure \ref{FAA} shows the energy dependence of ${\cal F}$ and ${\cal C}$
for polygonal structures from 3 to 6 vertices for a representative set 
of parameters. As mentioned above, the energies for which ${\cal F}$
vanishes are the physical eigenenergies of the polygon. We considered  
a representative energy interval
above the threshold $\varepsilon_1=\hbar^2\pi^2/(2md^2)$.
The energy $\varepsilon_1$ corresponds to the first transverse mode
of the 2D nanowire of width $d$ from which 
our 1D model derives (see \ref{apA}). 

From Fig.\ \ref{FAA} we notice that the polygon spectra are characterized by a sequence 
of doublets, with small energy splittings and with vanishing circulation ${\cal C}$.
In addition to the doublets, there are a varying 
number of single-energy modes (singlets) with finite ${\cal C}$ lying in between doublets.
For $N_v=3$, for instance, doublets around 5.02 and 5.13  with two intermediate singlets are seen in the upper left panel. Remarkably, doublets at the same
energies are present in polygons with odd and even
number of vertices. Polygons with an even number of vertices, however, possess additional doublets at intermediate energies.

Singlets in Fig.\ \ref{FAA} are characterized by having
${\cal C}\approx N_v \hbar p/m\equiv {\cal C}_0$. Notice, however, that
different values of the circulation are possible for each singlet energy.
Indeed, with the algorithm of \ref{apB} we find that a 
positive circulation ${\cal C}_0$
is obtained if requiring $a_l^{(i)}=1$
(for an arbitrary $i$) 
and negative circulation $-{\cal C}_0$ if
requiring $a_r^{(i)}=1$. This indicates that for each singlet 
the circulation can actually take any 
value within the range $-{\cal C}_0 < {\cal C} < {\cal C}_0$ by 
adequately superposing the solutions for positive and negative circulations.
We have explicitly checked this possibility from the numerical solutions.
This behavior manifests the connection between the symmetry breaking and
the ${\cal C}$-degeneracy of the singlets.   

A qualitative difference is seen when comparing left and right panels
of Fig.\ \ref{FAA}.
Polygons with an odd number of vertices have $N_v-1$ singlets in between 
vanishing-circulation doublets, while polygons with an even number of vertices only have $N_v/2-1$. 
This explicitly shows that by knowing the spectrum one may characterize the polygonal 
structure.
For instance, a spectrum consisting of a sequence of mode doublets
with three intermediate singlets would correspond to an octagon.

For the spectra with an even number of intermediate singlets a confusion 
might arise between the polygons with $N_v$ and $2N_v$ vertices. 
For instance,
both the triangle and the hexagon in Fig.\ \ref{FAA} have two intermediate singlets.
Still, one may differentiate the two situations with the discussion on doublets 
of next subsection. In essence, for odd-$N_v$ polygons all doublets are similar in that each
polygon side contains an even number of half wavelengths. 
On the contrary, for even-$N_v$ polygons consecutive doublets  alternate from an even to an odd number of half wavelengths.
As shown below, this difference can be seen counting the number of density maxima
on a side of the polygon.

\subsection{Doublets}

The states with vanishing circulation are characterized by having
$|a|=|b|$ on each side of the polygon. Besides vanishing circulation,
doublets have a density that remains
invariant by a translation from one polygon side to the next one
(Figs.\ \ref{FB} and \ref{FC}). This happens because the scattering amplitudes 
of successive vertices fulfill either $a^{(i)}=a^{(i+1)}$
or $a^{(i)}=-a^{(i+1)}$. At the same time, for a given 
vertex the coefficients also fulfill $a_l =\pm a_r$. 
Therefore, these
vanishing-circulation states can be classified in four types 
as summarized in Tab.\ \ref{tab1}. 

Table \ref{tab1} also gives the secular equation determining the 
wave number $p$ and the energy of the eigenmode
for each type of mode.
Notice that solutions of types I and III are associated to states
with maximal density on each vertex, while types II and IV have
minimal density on the vertex, cf.\ Figs.\ \ref{FB} and \ref{FC}. 
The split pair forming a doublet are therefore composed
by solutions of types (I,II) and (III,IV).
It is also 
worth  stressing that solutions of type III and IV are not allowed
in polygons with an odd number of sides. The reason is most 
easily understood assuming $r\approx 0$ and $t\approx 1$. In this
case, the secular equation for types III and IV 
simplifies to $e^{ipl}=-1$; that is $\ell = (2n+1)\lambda/2$
and
an odd number of half wavelengths should fit in a polygon side.
For odd values of $N_v$
this is not compatible with the additional condition that the full 
perimeter should contain and integer number $n'$ of wavelengths, $N_v \ell = n' \lambda$.

\begin{table}[t]
\centering
\caption{Different types of modes in vanishing-circulation doublets.
The pairs forming doublets are (I,II) and (III,IV). 
}
\begin{tabular}{cccc}
\hline\hline
type & left-right & successive vertices & equation \\
\hline
I \rule{0cm}{0.4cm} & $a_l=a_r$ & $a^{(i)}=a^{(i+1)}$ & $t+r=e^{-ip\ell}$\\
II \rule{0cm}{0.4cm} & $a_l=-a_r$ & $a^{(i)}=a^{(i+1)}$ & $t-r=e^{-ip\ell}$\\
III \rule{0cm}{0.4cm} & $a_l=a_r$ & $a^{(i)}=-a^{(i+1)}$ & $t+r=-e^{-ip\ell}$\\
IV \rule{0cm}{0.4cm} & $a_l=-a_r$ & $a^{(i)}=-a^{(i+1)}$ & $t-r=-e^{-ip\ell}$\\
\hline
\end{tabular}
\label{tab1}
\end{table}

Simple analytical approximations for the 
doublet splittings can be obtained from the secular equations
in Tab.\ \ref{tab1} assuming $t\approx 1$ and $r\approx r_0 e^{i\phi}$.
This leads to the conditions ($n=1,2,3\dots)$
\begin{equation}
\label{split}
\begin{array}{ll}
p\ell = 2 n\pi \mp r_0\sin\phi\;, & ({\rm I,II})\; ,\\
p\ell = (2 n+1)\pi \mp r_0\sin\phi\;,\quad &({\rm III,IV})\; .
\end{array} 
\end{equation}
For both kinds of pairs the splitting in wave numbers is $\Delta p=2r_0\sin\phi/\ell$,
with an associated energy splitting $\Delta=(\hbar^2/m)p\Delta p$.
Equation (\ref{split}) clearly shows that in (I,II) doublets, the solution of type
I decreases its energy due to reflection while that of type II increases, and analogously for (III,IV)
doublets. We recall that type I and III solutions are those with a maximal density on 
each vertex. Finally, we notice that as the energy increases the reflection coefficient is reduced, and so the 
doublet
splitting is also reduced, as hinted also in Fig.\ \ref{FAA}.

Summarizing, the energies of the zero-circulation
 doublets are approximately
 \begin{equation}
 \label{spec}
 E_{n\pm}= \varepsilon_1+\frac{\hbar^2\pi^2 n^2}{2m\ell^2}\pm \Delta\;,\quad
\!\!\!\!\!
\left\{
 \begin{array}{l}
 n=1,2,3,\dots ({\rm even}\, N_v)\; ,\\
 n=2,4,6,\dots ({\rm odd}\, N_v)\; .
 \end{array}
\right.
\end{equation} 
Physically, our result of Eq.\ (\ref{spec}) says that the arm length $\ell$ determines the separation
between successive doublets, while the reflection properties of the 
vertex determine the splitting $\Delta$ of the pair forming a doublet.

\subsection{Singlets}

\begin{figure}[t]
\centering
\resizebox{0.48\textwidth}{!}{
	\includegraphics{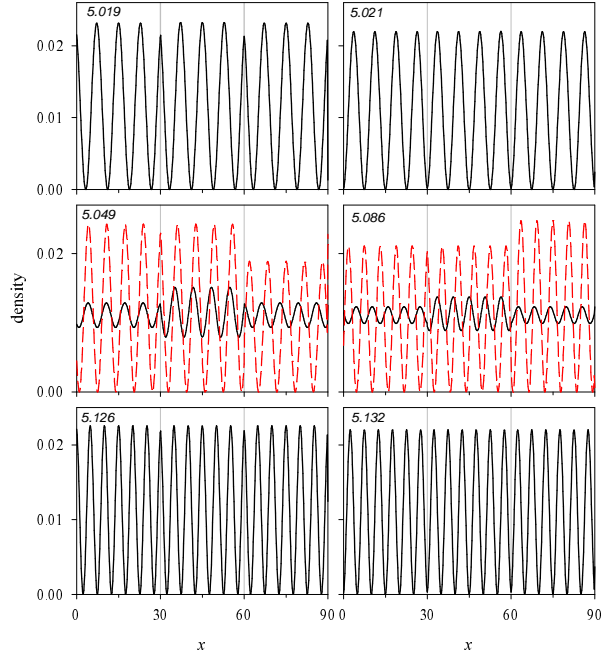}
}
\caption{
Densities for the triangle modes of Fig.\ \ref{FAA}. 
All distances are measured in units of $d$ (defined in \ref{apA}).
Upper and lower panels are for vanishing-circulation
doublets of type (I,II). Intermediate panels are singlets with maximal (solid) 
and vanishing (dashed) circulation. The energy of the mode (as in Fig.\ \ref{FAA}) is indicated in each panel. The position of vertices is given by the vertical lines. Rest of the parameters 
as in Fig.\ \ref{FAA}.  
}
\label{FB}
\end{figure}

\begin{figure}[t]
\centering
\resizebox{0.48\textwidth}{!}{
	\includegraphics{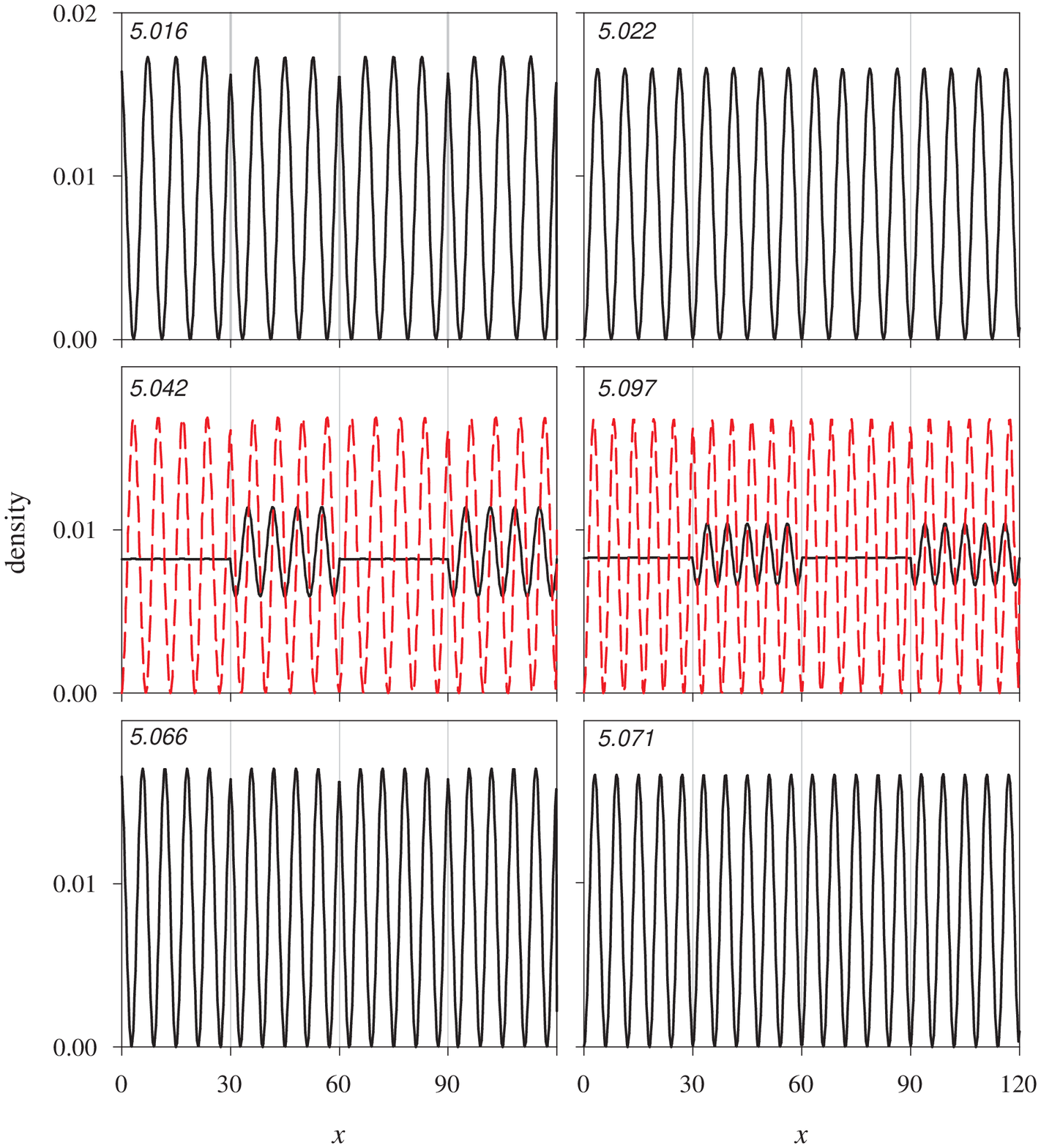}
}
\caption{Same as Fig.\ \ref{FB} for the square modes. 
Upper panels are the first pair [type (I,II)], while lower panels are the second pair [type (III,IV)]
in the results of Fig.\ \ref{FAA} for the square.
The two central panels 
are singlets with maximal (solid) and vanishing (dashed) circulation.}
\label{FC}
\end{figure}

The most remarkable feature of singlets is that 
the density is not equivalent on different sides of the polygon.
As mentioned above, for each singlet different solutions with
circulation ranging from $-{\cal C}_0$ to $+{\cal C}_0$,
where ${\cal C}_0\approx N_v\hbar p/m$, are possible.
The intermediate panels
of Figs.\ \ref{FB} (energies of 5.049 and 5.086) and in Fig\ \ref{FC} (5.042 and 5.097)
show the densities corresponding to singlets 
for the triangle 
and square.
Continuum and dashed lines are for states 
with maximal and vanishing circulation, respectively.
We notice that for states with non vanishing ${\cal C}$, the density 
for positive and negative circulations coincide.
For nonvanishing circulation the density does not vanish
at any point, but oscillates around a finite mean
value. In all singlets, different behaviors are seen on the sides of the 
polygon.

The singlet energies and their underlying ${\cal C}$-degeneracy can be 
easily understood in the limit of vanishing reflection. Indeed, when 
$r\approx 0$ and $t\approx 1$ the solutions to Eqs.\ (\ref{ivs})
and (\ref{bli}) decouple for positive and negative circulations.
The input coefficients for vertex $k$ read in this case
\begin{equation}
\label{eq8}
\begin{array}{cccc}
a_r^{(k)} &=& \alpha e^{ip\ell(k-1)}\;,\quad &({\cal C}<0)\; ,\\
a_l^{(k)} &=& \beta e^{-ip\ell(k-1)}\;,\quad &({\cal C}>0)\; ,
\end{array}
\end{equation}
where $\alpha$ and $\beta$ are arbitrary independent constants that can be 
fixed  by normalization. A maximal circulation state is obtained 
when either $\alpha$ or $\beta$ vanish. Arbitrary superpositions can 
be formed for $\alpha\ne0$ and $\beta\ne0$, including a vanishing circulation state 
for $\alpha=\beta$. The cyclic condition in Eq.\ (\ref{eq8}),
$e^{ip\ell N_v}=1$, leads to the energies
\begin{equation}
\label{eq9}
E_n=\varepsilon_1 + \frac{\hbar^2\pi^2 (2n)^2}{2m \ell^2N_v^2}\; ,\quad (n=1,2,3,\dots)\; .
\end{equation}
These energies coincide with a doublet condition 
[Eq.\ (\ref{spec}) for $\Delta=0$] each time $n$ is an integer number of times 
$N_v$. It can be easily verified that Eq.\ (\ref{eq9}) implies that for 
odd-$N_v$ polygons there are $N_v-1$ singlets between 
two successive doublets and that for even-$N_v$ polygons
there are $N_v/2-1$, in agreement with the results of Fig.\ \ref{FAA}

\section{Conclusions}

A model of closed polygons made of 1D quantum wires has been presented where each 
vertex is described as a scatterer. The vertex transmission and reflection coefficients
have been described with the model of circular bends in 2D wires.
The polygon spectra are characterized by a sequence of doublets, with a 
small energy splitting, and a typical number of singlets lying
in between doublets. Odd-$N_v$ polygons have $N_v-1$ singlets while
even $N_v$ have $N_v/2-1$ singlets in between two doublets.
The doublet splittings are caused by the reflection
on vertices. Doublets have a vanishing circulation and singlets can have 
circulations ranging from a negative to positive characteristic values.
Approximate analytical expressions for the singlet and doublet 
energies have been given and the corresponding densities discussed.
Our results  explicitly show how the polygon characteristics can be inferred from the 
spectrum.

\section*{Acknowledgement}
C.E. gratefully acknowledges a SURF@IFISC fellowship.
This work was funded by MINECO-Spain (grant FIS2011-23526),
CAIB-Spain (Conselleria d'Educaci\'o, Cultura i Universitats) and 
FEDER.

\appendix

\section{Scattering by a 2D circular bend}
\label{apA}

\begin{figure}[t]
\centering
\resizebox{0.28\textwidth}{!}{
	\includegraphics{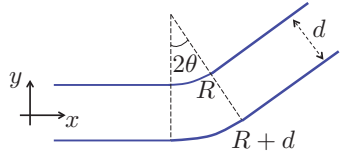}
}
\caption{Circular bend in a 2D nanowire.}
\label{bend}
\end{figure}

We consider a 2D nanowire with a circular bend of angle $2\theta$ and 
radius $R$, as sketched in Fig.\ \ref{bend}. 
Notice that $\theta$, defined as half the bend angle, is related in our model 
to the number of polygon vertices  by $\theta=\pi/N_v$.
The lateral extension of
the nanowire is $d$, setting the energy of the first transverse mode to
$\varepsilon_1=\hbar^2\pi^2/(2md^2)$. Following Ref.\ \cite{Sprung},
it is possible to derive an analytical expression for the reflection and transmission
coefficients of the circular bend in the single-mode limit.

Defining $a=\bar{R}\theta$, where $\bar{R}=\sqrt{R(R+d)}$ is an average radius,
the 2D scattering problem in the single mode limit is approximated by a 1D quantum well of 
width $2a$ and depth $V_0=-\hbar^2/(8 m \bar{R}^2)$. The analytical expressions 
of the $t$ and $r$ complex coefficients are given in terms of 
two wave numbers ($E>\varepsilon_1$)
\begin{eqnarray}
p &=& \sqrt{\frac{2m}{\hbar^2}(E-\varepsilon_1)}\; ,\\
q &=& \sqrt{\frac{2m}{\hbar^2}(E-\varepsilon_1-V_0)}\; .
\end{eqnarray}
Specifically, they read
\begin{eqnarray}
\label{tt}
t &=& \frac{4pq}{
 (p+q)^2 e^{2i(p-q)a}
-
 (p-q)^2e^{2i(p+q)a}
 }\; ,\\
 \label{rr}
r &=& -\frac{2i(p-q)(p+q)\sin(2qa)}{
 (p+q)^2 e^{2i(p-q)a}
-
 (p-q)^2e^{2i(p+q)a}
}\; .
\end{eqnarray}
Notice that, as expected, for large energies it is $p\approx q$ and
Eqs.\ (\ref{tt}) and (\ref{rr}) then yield $t\approx 1$ and $r\approx 0$.
We also stress that, due to symmetry, the transmission and reflection 
coefficients for left and right incidences are identical,
$t'=t$, $r'=r$, leading to the scattering matrix of Eq.\ (\ref{ivs}).

\section{Algorithm}
\label{apB}

Equations (\ref{ivs}), (\ref{bli}) and (\ref{bri}) are set up as a
homogoneous linear system
\begin{equation}
\label{hsys}
{\mathbf M} 
\left(
\begin{array}{c}
{\mathbf a}\\
{\mathbf b}
\end{array}
\right)  =0\; ,
\end{equation}
where we define the vector
\begin{equation}
{\mathbf a}^T \equiv
\left(
a_l^{(1)}
\dots
a_l^{(N_v)}
a_r^{(1)}
\dots
a_r^{(N_v)}
\right)\; ,
\end{equation}
 with an analogous definition for ${\mathbf b}$. 
 Nontrivial solutions of Eq.\ (\ref{hsys}) occur only for energies such that 
 the determinant of ${\mathbf M}$ vanishes, ${\rm det}({\mathbf M})=0$.

In practice,  we arbitrarily pick one of the 
 components of ${\mathbf a}$, say $a_\alpha^{(\beta)}$, and fix it to 1, transforming Eq.\  (\ref{hsys}) into the inhomogeneous problem
\begin{equation}
\label{isys}
{\mathbf M}' 
\left(
\begin{array}{c}
{\mathbf a}\\
{\mathbf b}
\end{array}
\right)  ={\mathbf R}\; ,
\end{equation}
where 
\begin{equation}
{\mathbf R}^T \equiv
\left(0\dots 0 1 0\dots 0\dots 0\right),
\end{equation} 
with the value 1 on the position of the chosen component $a_\alpha^{(\beta)}$.
Matrices ${\mathbf M}$ and ${\mathbf M}'$ are identical except for the 
row of the chosen component, which is diagonal for ${\mathbf M}'$.
Equation (\ref{isys}) can be easily solved by standard numerical routines.
The solution is then used to evaluate the following norm 
\begin{equation}
{\cal F} = {\rm norm}\left\{
{\mathbf M}
\left(
\begin{array}{c}
{\mathbf a}\\
{\mathbf b}
\end{array}
\right) 
\right\}\; .
\end{equation}

Obviously, when ${\cal F}=0$ the solution obtained from Eq.\ (\ref{isys})
is actually a solution of the original problem (\ref{hsys}).
The validity of the method is guaranteed by the linearity, 
for when a solution of (\ref{hsys}) exists it can always be scaled such that
a chosen component is equal to 1, provided only that it does not vanish. 
A simple energy scan of ${\cal F}$ will signal the 
position of the eigenenergies as the nodes of the function. 
We have checked that, as expected, the nodes are independent of the chosen component $a_\alpha^{(\beta)}$ and that 
they numerically coincide with the nodes of ${\rm det}({\mathbf M})=0$. 





\bibliographystyle{elsarticle-num}
\bibliography{polyrefs}

\begin{thebibliography}{10}
\expandafter\ifx\csname url\endcsname\relax
  \def\url#1{\texttt{#1}}\fi
\expandafter\ifx\csname urlprefix\endcsname\relax\def\urlprefix{URL }\fi
\expandafter\ifx\csname href\endcsname\relax
  \def\href#1#2{#2} \def\path#1{#1}\fi

\bibitem{Datta}
S.~Datta, Electronic transport in mesoscopic systems, Cambridge University
  Press, 1997.

\bibitem{Ferry}
D.~K. Ferry, G.~S. M., B.~Jonathan, Transport in nanostructures, Cambridge
  University Press, 2009.

\bibitem{Lent}
C.~S. Lent, Transmission through a bend in an electron waveguide, Appl. Phys.
  Lett. 56 (1990) 2554.

\bibitem{Sols}
F.~Sols, M.~Macucci, Circular bends in electron waveguides, Phys. Rev. B 41
  (1990) 11887--11891.

\bibitem{Sprung}
D.~W.~L. Sprung, H.~Wu, J.~Martorell, Understanding quantum wires with circular
  bends, J. Appl. Phys. 71 (1992) 515--517.

\bibitem{Wu}
H.~Wu, D.~W.~L. Sprung, J.~Martorell, Effective one-dimensional square well for
  two-dimensional quantum wires, Phys. Rev. B 45 (1992) 11960--11967.

\bibitem{Wu2}
H.~Wu, D.~W.~L. Sprung, J.~Martorell, Electronic properties of a quantum wire
  with arbitrary bending angle, J. Appl. Phys. 72 (1992) 151--154.

\bibitem{Wu3}
H.~Wu, D.~W.~L. Sprung, Theoretical study of multiple-bend quantum wires, Phys.
  Rev. B 47 (1993) 1500--1505.

\bibitem{Vacek}
K.~Vacek, A.~Okiji, H.~Kasai, Multichannel ballistic magnetotransport through
  quantum wires with double circular bends, Phys. Rev. B 47 (1993) 3695--3705.

\bibitem{Xu}
H.~Xu, Ballistic transport in quantum channels modulated with double-bend
  structures, Phys. Rev. B 47 (1993) 9537--9544.

\bibitem{Berko}
G.~Berkolaiko, P.~Kuchment, Introduction to quantum graphs, American
  Mathematical Society, 2012.

\end{thebibliography}







\end{document}